# The Discovery of Millisecond Pulsars: Don Backer and the Response to the Unexpected

P. B. Demorest and W. M. Goss
National Radio Astronomy Observatory, P.O. Box O, Socorro, NM 87801, USA
pdemores@nrao.edu, millergoss@gmail.com

## ABSTRACT

It has now been just over four decades since the first discovery of a millisecond spin period pulsar (MSP), B1937+21, by Don Backer and collaborators in late 1982. This finding of an entirely new class of astronomical object revolutionized pulsar astronomy and provided inspiration for novel scientific investigation for decades to come, continuing to the current day and beyond. Here we review the events leading to the discovery, based on archival material, personal correspondence, and first-hand accounts of several of the participants. We also briefly review the enormous impact that MSPs have had on physics and astronomy by highlighting major MSP-related science of the past 40 years.

## 1   INTRODUCTION

From 1975 to 1982, Don Backer had a quest that became a passion. After moving to Berkeley in 1976, he had numerous discussions with his colleagues, Leo Blitz and Stuart Vogel, about finding compact radio sources in the galactic plane, In the next year or two, Don continued conversations with Tony Readhead at Caltech about IPS (interplanetary scintillations of low galactic latitude compact radio sources.) In this era, Don accreted colleagues at conferences and observing trips to NRAO Green Bank and later the VLA. For those of us who knew Don well, he was the ultimate "networker" – perhaps the method of accreting contributing collaborators was different in the pre-email era? There were letters, telexes and telephone calls. As we will see, Don left a legacy of his scientific life that Goss was able to archive after Don's untimely death in 2010. During the process of the discovery of the 642 Hz pulsar (PSR B1937+21) in November 1982, Don had ensured that each member of the collaboration (Shri Kulkarni, Carl Heiles, Mike Davis and W. Miller Goss) would bring unique contributions to the endeavor.

### 1.1   Don Backer: A Brief Biography

Donald C. Backer was born in New Jersey in 1943. He received a BS in engineering physics from Cornell University in 1966 and a MSc from the University of Manchester in 1968 working at the Jodrell Bank radio telescope. In 1971, Backer received his PhD in astronomy from Cornell. His PhD thesis, titled "Radio Intensity Flucutations in Pulsars," was based on Arecibo data. He investigated pulsar pulse-to-pulse variations and scintillations, under the direction of Frank Drake. He discovered "nulling pulsars," pulsars that occasionally stopped pulsing. Immediately following his PhD, Backer moved to the NRAO in Charlottesville, as an NRAO postdoctoral fellow from 1971 to 1973, followed by a two-year appointment at NASA Goddard Space Flight Center from 1973 to 1975.

Backer's career blossomed in 1975 as he moved from NASA to the University of California, Berkeley, Radio Astronomy Laboratory. He became a professor in 1989, remaining on the Berkely faculty until his unexpected death on 25 July 2010. Further details of Don's remarkable scientific career were presented in a *Nature* obituary by Kulkarni (2010).

### 1.2   The Backer Archives at the NRAO

A key resource for the preparation of this article is an archive of Don's papers that is maintained by the NRAO. Since 1979, Goss maintained a partial archive of the 1982 discovery in his files, but some hand-written 1982 letters to Backer were not retained and further relevant material was lost during an international move in 1986.

In late October 2010, a few months after Don's death, a celebration of 40 years of the Westerbork Synthesis Radio Telescope was held at the Westerbork observatory site in Drenthe, the Netherlands. Goss gave a presentation on 22 October 2010: "*The Role of the WSRT in the Discovery of the First Millisecond Pulsar PSR 1937+21- In Memoriam Donald Charles Backer 9 November 1943-25 July 2010.*" Imke de Pater (Chair of Astronomy Department, University of California, Berkeley, whose PhD had been awarded at Leiden for research with the WSRT) was present. She invited Goss to visit Berkeley in early 2011 to help with the massive chore of making an inventory of Don's office at Campbell Hall on the Berkeley campus.

In January 2011, Goss was in Berkeley for a few days on the way to Sydney. Dan Werthimer, Geoff Bower, David DeBoer and others helped with an initial sorting: all UC Berkeley materials were sorted by local UC staff. On the other hand, Goss looked at other materials with the intention of transferal of relevant papers (Backer science projects, correspondence with non-UC colleagues and especially US radio astronomy committee reports) to the National Radio Astronomy Observatory, Associated Universities Inc. Archives (Charlottesville, Virginia, Ellen Bouton, Archivist). Goss was especially thrilled to find the file titled "W.M. Goss"! All missing letters, many handwritten, from 1979 to 1985 written by Goss were located. These included several key letters from the discovery year 1982 which had not been preserved by Goss in his files. All this material (Backer and Goss) is now preserved in the NRAO Archive.

## 2    ACCOUNT OF THE DISCOVERY

Here we recount the events leading up to the discovery of B1937+21 as a millisecond-spin-period radio pulsar. We start with a detailed timeline of the identification and investigation of the radio source 4C21.53 and then present additional details of the discovery observations at Arecibo in 1982.

### 2.1    Discovery timeline

In 1965, the 4C survey (Pilkington and Scott, 1965) provided the first detection of the 4C source 4C21.53. The properties were RA (1950) $19^h$ $38.1^m$ , Decl.(1950) 21 deg 34. min, S=2.9 Jy with errors of 0.6 and 4 arc min and about 25 per cent in flux density.

From 1967 to 1976 the Cambridge Cavendish Laboratory group of Hewish, Readhead, and collaborators carried out a program of observing interplanetary scintillation of radio sources. This program notably resulted in the original discovery of pulsars (Hewish et al., 1968). Scintillation observations of 4C21.53 were taken as part of a survey by Readhead (1972). A detailed account of the history of these observations and analyses is presented by Readhead (2024, JAHH, submitted).

In 1974, Readhead and Hewish (1974) determined that 4C21.53 was less than 1 arc sec in size by observing interplanetary scintillation of the source with the array at Cambridge.

In 1976, Duffett-Smith and Readhead (1976) used part of the 4C Telescope, at 151.5 MHz, and the existing 81.5 MHz data, to investigate interstellar scattering. They pointed out that the scintillation in 4C21.53 observed at 81.5 MHz by Readhead and Hewish (1974) is anomalous.

In Berkeley in 1977 and 1978 Backer and Stuart Vogel searched for compact radio sources to observe with VLBI in Cygnus in order to access the presence of interstellar scattering. In 1979 a key event occurred as Tony Readhead told Don Backer about a few low galactic latitude steep spectrum sources which showed IPS – interplanetary scintillators, in particular 4C21.53. Readhead pointed Don to a paper by Rickard and Cronyn (1979) from the University of Iowa, titled "Interstellar Scattering, the North Polar Spur, and a Possible New Class of Compact Galactic Radio Source." [1]

> Rickard and Cronyn wrote:
>
> The presence of so many small sources in the [galactic] plane in a strong scattering medium requires that the sources be nearby galactic sources- not extragalactic. Let us call these hypothetical sources scintars because, at present, their IPS activity is the one observational characteristic which distinguishes them from other galactic sources. **Scintars** are compact radio sources with typical fluxes less that 20 Jy [at low radio frequencies] which for intrinsic angular sizes (less that 0.5 arc sec), imply brightness temperatures in excess of $10^{10}$ K, i.e. they must be non-thermal. If we were to single out one source from this last as worthy of special mention it would be 4C 21.53.[42 objects were scintar candidates.][2]

Backer began to plan a campaign to follow up on the scintar suggestion. On 10 January 1979 he wrote a two-page hand-written letter to Mike Davis at Arecibo:

> Most objects in the galactic plane that [Readhead] studied did not scintillate … likely related to angular broadening in interstellar scintillation (ISS). Perhaps this object [4C21.53] is seen through a "hole" in the interstellar medium, perhaps it is a nearby pulsar.
>
> … Hulse and Taylor surveyed this region [an earlier Arecibo observation] down to 1.5 mJy with P greater than 60 ms and dispersion measures less than 1280 pc/cm$^3$. Perhaps the pulse is smeared by ISS: we would expect the time of smearing to be much greater than the period. But we know that the source size is not smeared beyond about one arc second at 81 MHz – otherwise no IPS. For a hypothetical distance at 5 kpc [a remarkable good guess for the HI distance determined later to be 5 kpc! – see below], the time of smearing is then roughly less than 10 ms, if it is a pulsar.
>
> [Backer continued] the following questions arise:
>
> 1. Is this a kHz pulsar? A recording at S band [at Arecibo] would be interesting
> 2. Does the meter wave scintillator have a high frequency halo [related to the cm source detected at Parkes at 11 cm]? A drift scan at S band would tell.
> 3. What is the spectrum between 100 and 600 MHz? Not so easy but a search for variation at 430 MHz [at Arecibo] would be interesting,

Davis followed this suggestion on 17 January 1979. He carried out a brief Arecibo search for pulses, searching for periods greater than 55 ms for 4C21.53. This search was doomed due to imprecise coordinates at this time [errors of 5 to 10 arcmin].

On 7 March 1979, Don Backer and colleagues Stuart Vogel, David Cudaback and John Middleditch used the Owens Valley Radio Observatory (Caltech) 90-foot telescope for a pulsar search of 4C21.53 at 600 MHz (primary beam about 1.2deg). The search was sensitive to periods greater than 20 ms. Due to extensive radio frequency interference the observations were corrupted.

---

[1] Backer (1984) described the impact of the new scintillation data of 1974-1979: "The source 4C21.53 came to our attention since it displayed strong interplanetary scintillations despite its low galactic latitude. Pulsar observations had indicated that at low galactic latitudes interstellar scattering (ISS) would suppress the IPS modulation. While identification of 4C21.53 with a pulsar would have explained its peculiar properties, no known pulsar was within the errors of the 4C21.53 object."

[2] Starting in 1984 Kulkarni and Goss have shared a whimsical aversion to the term "scintars", based on the failure (see below) to detect any pulsars associated with such sources. PSR B1937+21 remains the only detected pulsar based on this approach.

On 30 March 1979, Don Backer submitted a paper to *Astronomy and Astrophysics Letters* (Editor Stuart Pottash, University of Groningen, The Netherlands), "Possibility of a New Pulsar-Supernova Remnant Pair". Backer wrote in his retrospective article (Backer, 1984) as he described events in 1979:

> While searching the literature for references to 4C21.53 in January 1979, I found a source, 1937+215, located 30 s west (7 arcmin) of the 4C21.53 position in several published catalogues. This source could be identified with 4C21.53 if the 4C position was in error by one lobe. Lobe errors had been recognized in the 4C Catalogue occur with a frequency of about 3 per cent (Backer et al. 1970). The difficulty with connecting 1937+215 and the IPS source was that the spectra were very different. Furthermore, the large size, 60 arcsec, for 1937+215 in the 5 GHz catalogue of Altenhoff et al. (1979) confounded the lobe-error hypothesis since the IPS object was necessarily smaller than 1 arcsec. My initial synthesis of these observations was that the steep-spectrum, IPS object was a pulsar co-located with a faint (1 Jy), extended supernova remnant, 1937+215. Curiously no pulsar had been detected in this region in the very sensitive Arecibo survey (Hulse and Taylor 1974) … A simple calculation demonstrated that ISS could only smear pulsations for periods of order 10 ms or less. The Arecibo survey, and most others, were not sensitive to periods below 60 ms. A report on the hypothesis that 1937+215 was a young pulsar-supernova pair similar to the Crab nebula and its pulsar – published prior to the discovery of pulsars (Hewish & Okoye 1965) – was returned from a journal with the referee's comment as "**too speculative**" [our emphasis].

There is more to this story: Goss was the referee for the Astronomy and Astrophysics letter of March 1979. Backer's claim in 1984 about the fate of the paper was not entirely correct. Based on the Backer draft manuscript from 21 March 1979 ("Possibility of a New Pulsar-Supernova Remnant Pair"), 3 typed pages, two tables and two figures, we can now summarize the complex series of events that led to his conclusions in the submitted manuscript:

The IPS of 4C 21.53 reported by Readhead and Hewish indicated an upper limit of about one arc sec for this low galactic latitude (-0.3 deg) radio source with a steep spectrum (see Fig. 6 from Backer, 1984). Backer found a number of determinations of two sources close to the 4C position but displaced to the west of the 4C position by about 8 arc min, 1937+215. There was an extended flat spectrum source, detected at 6 cm (140-foot Green Bank) and 11 cm (Parkes) with a size of about one arc min. In addition, a low frequency compact source had been detected at 80 MHz (with the Culgoora Radioheliograph, Slee and Higgins, 1975) 8 arc min to the west of the 4C position, confirming the lobe shift suggested by Backer. The latter component was slightly displaced to the south of the extended component, later known to be a HII region G57.6-0.3, at a distance of 10.7 kpc compared to the HI-determined distance for the pulsar at 5 kpc (Heiles et al., 1983).

Goss's referee report was given to Stuart Pottash, the editor of Astronomy and Astrophysics Letters on about 15 April 1979 (Goss was a colleague of Pottash at the University of Groningen, the Netherlands, Kapteyn Astronomical Institute). Goss's main objection to the Backer paper was the suggested parallel between the 4C21.53 components and the Crab Nebula, an extended supernova remnant from SN 1054 with an associated fast pulsar (33 msec). Backer wrote:

> This summary of the extant observations of 4C21.53 … is reminiscent of the pre-pulsar discovery of the compact source in the midst of the Crab Nebula supernova remnant….

However, Backer's conclusion of the 1979 manuscript is prescient:

> Based on the observations described above, the probability that 4C21.53 is a pulsar is very high. The negative results of Hulse and Taylor [high sensitivty survey at Arecibo at 430 MHz, 1974] may be attributed to … a short period (less than 10 ms? [his question mark]), which could be smeared by interstellar scattering consistent with the interplanetary scintillation observed at 81 MHz.

Pottash wrote to Backer on 20 April 1979 with the report of the referee:

> As you can see from the [referee's] report, he feels the paper is interesting and worth publishing as a Letter, but in the present should not be published because the arguments are confused, there is too much speculation, and the paper is poorly written.[3] I therefore request that you completely rewrite the paper according to the suggestions of the referee. I look forward to receiving a revision as soon as possible.

Backer's assertion that the rejection was due to the referee's assertion "too speculative" is not justified. In circa 2005, Don Backer told Goss in a phone conversation that the 1979 paper was indeed "poorly written."

In the period 1979 to 1981, Backer's interests turned to another component of 4C21.53, a component 14 arcmin displaced to the east of the main component. Erickson had pointed this out to Backer in a personal communication in 1979. For a short period in late 1979 or early 1980, Backer thought that the E component might be the relevant source since Erickson pointed out that the new component was also a scintillating source. But almost immediately Backer (1984) realized this was not decisive:

> In 1981 interest in the western source (1937+215) was rekindled by a report from Erickson [a short report in an abstract from an American Astronomical Society meeting in 1980] that 34-MHz data from the Clark Lake Radio Observatory indicated IPS [interplanetary scintillations] at the positions of both the E and the W components of 4C21.53. The pulsar-supernova hypothesis was resurrected.

However, this proposal became quicky unlikely since Backer quickly found (VLA data at 6 cm) in 1981 that the E source was a double separated N-S by 0.8 arc sec. Thus, this component was most likely extragalactic.

At the same time, Bill Erickson was making rapid progress in determining the nature of 4C21.53. He and Backer (plus collaborators) were trying to determine the source positions and spectral index distributions. Erickson published two papers in 1983, the first in Astrophysical Journal Letters in the January 1 issue (Erickson, 1983), a few weeks after the publication of the Backer et al millisecond pulsar detection in Nature(Backer et al., 1982a).[4] Erickson's paper (title "What is 4C 21.53?") consisted of a 21 cm VLA image with a resolution of 1.4 by 2.9 arc. The extended source (now known to be an HII region, G57.6-0.3) was proposed to be a possible flat spectrum supernova remnant associated with a pulsar, reminiscent of the Crab Nebula. However, Erickson qualified this suggestion: "The final possibility is that the extended component of 4C 21.53 an HII region. While the compact component is unrelated, steep-spectrum source, probably a strong, highly dispersed pulsar." This suggestion turned out to be correct.[5]

On 18 March 1981, Backer wrote a letter to Goss at the Kapteyn Lab in Groningen. Backer inquired if a Westerbork Synthesis Radio Telescope (WSRT) 49-cm image of 4C21.53 could be obtained (resolution about 0.5 arcmin), a newly commissioned wavelength. The idea was to observe with enough resolution to sort out the various components of 4C21.53, the confusing source to the east and the sources to the west: (1) the flat spectrum extended source and the compact component.

---

[3] Examples were the first lines of the discussion section written in an unconventional telegraphic style: "(1) Is 4C21.53C (compact) a pulsar? Period? Spin Down? Dispersion measure? …. Can we obtain an HI distance estimate for 4C21.53 (extended)?"

[4] The second publication was Rickard et al. (1983). Forty scintar candidates were observed with the VLA at 20 cm. Two-thirds of sources were double sources (extragalactic radio galaxies) suggesting that scattering at meter wavelengths along the galactic plane was less than had been predicted by the originators of the scintar classifications, Rickard and Cronyn (1979).

[5] At the 1984 Green Bank Millisecond Pulsar Workshop, Backer pointed out this amazing coincidence regarding 4C21.53: "The 4C instrument would have bypassed 1937+214 had it not been for the coincidental arrangement of three unrelated objects [radio sources] of comparable strength in two of the 4C interferometer lobes. Any one of the sources would not have been above the 4C intensity threshold. If 1937+214 had been isolated on the sky, we might not be here today!"

On 9 April 1981, Goss replied to Backer. He agreed to Don's suggestion to propose a WSRT observation. Backer then drafted a WSRT proposal to image 4C21.52 at 49-cm wavelength. Goss edited the text, then submitted it to the WSRT committee in September 1981 for the next WSRT observing cycle. Goss was Chair of the Proposal selection committee (the PC) at this time.

The PC met on 9 October 1981. In addition to Goss, the committee included Jan Oort (Leiden), John Baldwin (Cambridge), and Tony Willis (NFRA) among others. In the discussion, Goss, as chair of the PC, asked that Willis led the deliberation. Goss did remain in the meeting but made no comments on his own proposal. In the modern era, clearly Goss would have left the room during the discussion! At the WSRT in 1981, this was not done.

Baldwin made the comment that it was an "ok project" but he was doubtful the proposed observations of 4C21.53 would lead to a decisive result. As was often the case, Oort was then asked for his assessment; his remarks carried the day: "speculative proposal, but we should take a chance."[6] The proposal was accepted. Goss then resumed his chairmanship of the meeting, as additional proposals were considered. This 4C21.53 proposal was granted 12 hours at 49 cm.

On 15 January 1982, 4C21.53 was observed for 12 hours at 49 cm. Both components of 4C21.53 were covered by the primary beam of about 1.2 deg.

On 1 March 1982, Goss posted a letter to Backer about the 49 cm observations. Only Stokes I images were available; Goss had tried with no success to make images in all four Stokes parameters. The software for Stokes Q, U and V was not functional; this would be available only by September 1982. The 49 cm I image is shown from this date in Figure 1 (from the Goss archive). As Don wrote (1984): "This was the first hard evidence for the location of the IPS source in the western component. The intensity, 0.13 Jy, confirmed the steep spectrum seen at decametric wavelengths." The 0.13 Jy source was offset from the flat spectrum, extended source to the north by almost 3 arc min.

At this time, a short Arecibo observation at the WSRT position of the compact source was carried out by Val Boriakoff. Observations were made, searching for periods >4 ms. The results were inconclusive.

On 24 March 1982, Goss was visiting the Very Large Array of the National Radio Astronomy Observatory in New Mexico (USA). He had a one-hour phone call with Backer about the status of the WSRT data, especially the missing polarization image. Backer made detailed notes which were found in his Berkeley files after his death. Examples of items included were the draft proposal for WSRT data; the completion of the January 1982 49-cm WSRT data; and discussion of a VLA C-configuration proposal to observe the 4C21.53 field.

Goss and Backer had submitted a 21-cm proposal for a WSRT observation of 4C21.53, a follow-up to the successful 49-cm earlier observation of 15 January 1982. At this meeting of the WSRT PC, (4 June 1982), in contrast to the previous proposal, John Baldwin was favourable after seeing the 49-cm images. The new 21-cm observations occurred at Westerbork on 8 August 1982.

In August 1982, Backer and Goss attended the International Astronomical Union (IAU) at Patras, Greece. The two colleagues met for numerous discussions. The relaxing discussions were usually outside the conference in the warm summer surroundings on a lawn, frequently with Don's teen age son David Backer lying in the grass reading a book. We discussed at length the significance of the 49-cm image with the compact radio source and the northern extended source. We suspected a pulsar was associated with the southern point source; however, a number of uncertainties remained. Don and Shri

---

[6] In this era, if Oort were present at the PC, he was frequently asked to provide an opinion on controversial topics. His opinion was usually accepted.

Kulkarni – a graduate student of Prof. Carl Heiles at Berkeley (University of California) – were already planning an observing campaign for a proposed pulsar search at Arecibo, later in 1982.

By September 1982, the polarization software for WSRT data in Groningen had become operational. Goss sent Backer a telex to Backer on 13 September with amazing news: the point source in 4C21.53 was 28 percent linearly polarized at 49 cm, likely a pulsar. Goss wrote in his telex: "It looks like [the] steep source must be a pulsar." The point source had a steep spectral index with 130 mJy flux density at 49cm, compared to only 17 mJy at 20 cm (telex of 30 September 1982). At 20cm the compact source showed 15 percent polarization. A few days later on 5 October, Goss posted the two contour images to Backer (at 20 and 49 cm). Don immediately replied by telex: "Exciting result. We have time for pulsar search [Arecibo] on 25 September." The 20 cm image with a beam of 13.3 by 37.0 arc sec is shown in Figure 2 (from the Goss archive); again the pulsar candidate is the compact source to the south.

In the meantime, on 25 September 1982, Shri Kulkarni carried out an irregularly scheduled observation at Arecibo at the position of the compact steep spectrum component detected at 20 cm with the WSRT. A detailed first-hand account of this observation is given by Kulkarni (2024, JAHH, submitted). Two harmonics of a 1.558-ms period pulsar were detected for 3 of the 7 min successive observations. The next day no pulses were detected at 1400 or at the higher frequency of 2380 MHz. The situation was encouraging but remained uncertain.

On 7 October 1982, an official proposal had been given to Don Campbell (Director of Arecibo Observatory) with a request for four hours a day for data of 4C21.53 on 4-8 November, followed later on 10 and 12 November 1982. The pulsar search was being intensified, likely encouraged by the promising results from late September.

On 3 November 1982, Backer arrived at Arecibo Observatory (AO) in Puerto Rico, followed by discussions with the AO collaborator Mike Davis. Observations began the next day. On 5 November, Carl Heiles was present as Shri Kulkarni arrived. On 6 November 1982 a search was carried out looking for deep ISS based on the possibility that the compact object was small but not pulsating. At 21 cm, the frequency and time correlations were 2 MHz and 5 min, indicating a small angular size; previously only pulsars had exhibited ISS. The modulation bandwidth and time scale were consistent with the pulse detection in late September.

On 7 November 1982, the pulsar was detected with an interpulse and pulse of comparable intensities (Figure 3, from the NRAO/AUI Archives, Backer Papers).

On this date, Heiles called Goss with an update in the Netherlands at 3 am Middle European Time (8 November in Europe) with a summary of the new observations at Arecibo. Goss attempted to make detailed notes with a befuddled brain in the middle of the night! The next morning, he found that most of his text was a collection of confused notes. Later that day the pulsar was observed again at Arecibo with the same pulse rate of 642 Hz.

On 9 November 1982, Don Backer wrote in his notebook that he was ill – "in bed" – and that he needed to "visit the doctor for infection." In his sick bed, he was busy calculating the dispersion measure of the pulsar and an inferred distance based on the modest dispersion measure of 71.2 pc per $cm^3$. He suggested an inferred distance of 2.5 kpc based on an assumed (and uncertain) electron density along the line of sight of 0.03 electrons per $cm^3$. He also calculated a rate of energy loss and the age of the pulsar.

Finally, Don was planned to contact Joe Taylor at Princeton to begin accurate timing of the pulsar PSR 1937+21. Don wrote in his notebook "BINGO" in large letters as he privately celebrated his amazing success.

On 12 November 1982, the IAU circular (No 3743; copied here as Figure 4) published with details of the PSR 1937+21 detection (Backer et al., 1982b). The reported period derivative of $3 \times 10^{-14}$ sec per sec had been based on the apparent change in the period from September to November; this data turned out to be vastly in error, too large by a factor a huge four orders of magnitude. The cause was a sampling error in the first observations from September 1982.[7]

On 19 November 1982, Goss a sent a telex to Radhakrishnan (Rad) at the Raman Research Institute in Bangalore, India. "In September and last week, pulses have been found at Arecibo…The period is 1.557708 ms. The period derivative is quite uncertain." The latter warning was to have major consequences. Rad and G Srinivasan began work immediately on a paper. Their interpretation proposed that PSR 1937+21 was an old pulsar spun-up by mass transfer in a binary system, a **recycled pulsar** (see below).

On 22 November 1982, Backer (with co-authors Kulkarni, Heiles, Davis, and Goss) submitted the *Nature* paper "A Millisecond Pulsar" (Backer et al., 1982a, accepted 25 November 1982, with 713 citations to date). The paper also included a discussion of the extended source about 3 arc min to the north of the pulsar. A H166 alpha radio recombination was detected at Arecibo indicating that the flat spectrum source was an HII region. Later HI Arecibo observations indicated the two were not associated, with the HII region being at a distance of about 11 kpc, with the pulsar at a distance of 5 kpc. The assertion by Backer et al (in the Nature paper) that the HII region and the PSR might be associated, based on the similar locations on the sky and an implied assertion that the exciting star of the HII region might be the escaped binary companion of the pulsar was not confirmed.

On 23 November 1982, Don Backer presented a colloquium at Berkeley about the new pulsar at 4 pm in 3 Le Conte Hall: "A Millisecond Pulsar"

26 November 1982, Backer contacted Joe Taylor at Arecibo Observatory for accurate timing of the pulsar at 1408 MHz. It was necessary to sort out the uncertain period derivative. The results were published in *Nature* on 27 January 1983 (Backer et al., 1983). The period derivate was found to be $1.2 \times 10^{-19}$ sec per sec with a characteristic age of $4 \times 10^8$ yr.

On 16 December 1982, the discovery paper was published in *Nature* (Backer et al., 1982a). The cover has a subtitle "A Millisecond Pulsar" along with an image of two "gigantic" mice who had profited from growth hormone genes.

David Helfand wrote a News and Views text "A Superfast New Pulsar" appearing in the same *Nature* issue as the discovery (page 573), including several notable quotes:

1. "[It had been assumed, based on the Crab nebula and the associated 33 ms period pulsar], that there no younger and faster pulsars to be found. The assumption, that all fast pulsars are young, may have been the fatal flaw in the argument."

2. "The rotation rate of PSR 1937+21 is within a factor of about 3 of centrifugal break-up. The star's surface is moving at about 20 per cent of the speed of light and the rotational kinetic energy of the object is nearly $10^{52}$ ergs, comparable with the entire energy of a supernova explosion."

3. "At the moment, though, 4C21.53 serves as a reminder of the physical Universe in which each unexplored corner of parameter space holds a new mystery for the astrophysicist to ponder."

## 3 AFTER THE DISCOVERY

Immediately after the discovery of B1937+21 some of the most pressing questions were: On the theoretical astrophysics side, explaining the nature of the source and how it had come to be spinning at such an extreme rate; and on the observational side to search for other similar objects.

---

[7] The Nature publication of 16 December 1982 quoted an upper limit of $10^{-14}$ sec per sec for the period derivative.

## 3.1 Explaining Millisecond Pulsars

Radhakrishnan and Srinivasan (1982) published in Current Science, "On the Origin of the Recently Discovered Ultra-Rapid Pulsar." Based on the news from Goss from 13 November, the authors in Bangalore, India (Raman Research Institute) began work immediately on a short paper which was published on 5 December 1982. Due to the absence of a detected supernova remnant or a detected supernova, the authors suggested that the period derivative must be less than $10^{-19}$ sec per sec with a magnetic field of order $10^8$ G. "It appears to us much more reasonable to assume that both the short period and the very low field have a common cause which is relatively rare." The spin-up occurred in a long-lived mass transfer binary – rather than being young, this is a **recycled** old pulsar. The authors continued:

> [The new pulsar associated with 4C21.53 acquired] its rapid rotation rate as a neutron star in an accreting binary system in which the evolution time for the secondary was (1 – 2) x $10^7$ y. During this time, its initial field decayed to $10^{8-9}$ G and led to an equilibrium period of about 1.5 ms during accretion. The disruption of the binary system by the second explosion in it took place at least several thousands of years ago, so that the remnant has been completely dispersed in the ISM. The companion neutron star would have moved away in distance by an amount proportional to this unknown time and may or may not be observable,

On 23 December 1982 a Nature article by Alpar et al. (1982), "A New Class of Radio Pulsars" also proposed the recycling scenario. The paper appeared the following week as the Backer Nature discovery paper. In this "News and Views" of 16 December (see above), David Helfand had already anticipated the Alpar et al paper:

> Backer et al suggest [in their Nature paper] that the pulsar and the nearby diffuse source (now known to be an HII region) may be associated, with the latter excited by an escaped binary companion of the fast pulsar. From the separation of the two sources and using a typical pulsar velocity, they infer an age of the pulsar of about $10^4$ years. The case is far from compelling, however. A Alpar, M Ruderman and J Shaham of Columbia [colleagues of Helfand], together with A Cheng from Rutgers University, have proposed a scenario in which the neutron star is billions of years old and was reactivated as a pulsar after having been spun up by an accretion from a companion. [He described the paper that was to appear in the following week's Nature.] [Alpar et al] suggest the source is the remnant of a galactic bulge [low-mass] X-ray binary system. The standard scenario for bulge sources consists of an evolving low-mass star spilling matter onto a neutron star with a very low magnetic field… [A large transfer of angular momentum is transferred over a period of $10^8$ to $10^9$ years], spinning it up to millisecond periods. The ultimate spin rate is a function principally of the field strength and the mass transfer rate. Once the accretion stops as a result of the dissipation or ejection of the companion [star], the pulsar will begin to spin down again. But with its low field, the spindown time will be very long, preserving a population of old rapidly spinning pulsars.

In a review article by van den Heuvel (2017) in the Journal of Astrophysics and Astronomy, "Formation of Double Neutron Stars, Millisecond Pulsars and Double Black Holes," the author provided a fascinating history of his reaction in 1982 to the papers by Radhakrishnan and Srinivasan, Alpar et al and somewhat later by Fabian et al. (1983):

> Immediately after this discovery [of the 642 Hz pulsar PSR 1937+21], Radhakrishnan and Srinivasan (1982) and Alpar et al. (1982), and somewhat later Fabian et al. (1983), independently put forward the idea that, like the Hulse-Taylor binary pulsar PSR B1913+16, this pulsar has been recycled in a binary system. This time: not in a High-Mass X-ray Binary, where the accretion and spin-up phase lasts relatively short (at most $\sim 10^6$ years), but in a Low-Mass X-ray Binary (LMXB), where it may last $10^8$ to $10^9$ years, such that a very large amount of mass and angular momentum can be fed to the neutron star. They extended here the recycling idea by a factor 30, relative to the 59 msec pulse period of PSR B1913+16. And on top of that they had to assume that somehow the companion star in the progenitor binary, that had fed the mass and angular to this neutron star, had disappeared. When I read these first two papers, my first reaction was: "this is ridiculous, as it takes the recycling idea far out of its range of applicability". However, within a year, Radhakrishnan and Srinivasan, Alpar and colleagues and Fabian and colleagues, were proven completely right, thanks to the discovery by Boriakoff et al. (1983) of the second millisecond pulsar, which indeed is in a binary system, with a low-mass helium white dwarf as a companion star.

# 4 SCIENTIFIC LEGACY OF MILLISECOND PULSARS

The discovery of millisecond pulsars has had scientific implications reaching far beyond the immediate goals of characterizing the MSPs themselves and explaining their nature and origins. These primarily result from the incredible utility of MSPs as a kind of naturally-occurring "signal generator" that can probe the intervening space and material between the pulsar and the Earth. Two features of the MSP signals in particular are worth pointing out: First is the high spin rate which results in very sharp pulse shapes, as low as $\sim 50-100$ $\mu$s in the best cases; this directly translates into orders-of-magnitude more precise pulse arrival time measurements than had been possible with the "slow" (also known as "canonical") pulsars previously known. The second is that MSPs have extremely stable and predictable spin rates. In the best cases the observed spin frequencies deviate from a simple spin-down model by less than one part in $10^{15}$ over time periods of years to decades. This can again be compared to canonical pulsars where a typical value is roughly three orders of magnitude larger (Shannon and Cordes, 2010); very young pulsars such as the Crab are even less stable.

The combination of high measurement precision and clock-like stability have been used in the four decades since the original MSP discovery to make a number of unique measurements that otherwise would have never been possible in an astrophysical context. Here we will highlight a few of the most impactful results. We start with measurements of binary MSP systems that resulted in powerful tests of theories of gravitation and nuclear physics. This is followed by an account of ongoing attempts to use MSP timing data to detect very low frequency gravitational waves. While this section highlights many notable results, this is not intended to be a complete review of all MSP-related science over the past four decades (see review articles by Lorimer, 2008; Manchester, 2017).

## 4.1 Binary Systems as Physics Laboratories

The use of pulsars as probes of physics unrelated to the actual radio emission mechanism itself pre-dates the discovery of MSPs, although they eventually will become the primary target for this sort of research. Most famously, the first pulsar in a binary system, PSR B1913+16 was discovered in 1975 as part of an Arecibo pulsar survey (Hulse and Taylor, 1975). This system[8] already hinted at the formation pathway for MSPs. With a spin period of 59 ms, it was the second-fastest spinning pulsar after the Crab. In a 7.7-hour orbit, the companion was of comparable mass; as noted in the discovery paper "the mass ratio cannot be very different from unity … We conclude that the companion must be a compact object, probably a neutron star or black hole." Further timing from Arecibo eventually determined the masses to be nearly identical (both about 1.4 solar mass), and detected the decay of the orbit due to emission of gravitational radiation in exact agreement with the predictions of general relativity (Taylor and Weisberg, 1982). This provided the first – and for decades only – experimental evidence for the existence of gravitational waves, resulting in a Nobel Prize for the discoverers. A similar but even more extreme example of a such a system was found in 2003, in a 2.4-hour orbit with both stars visible as radio pulsars with spin periods of 22 ms and 2.8 s (Lyne et al., 2004). This system provides even stronger constraints on the correctness of general relativity (Kramer et al., 2006).

As additional MSPs were discovered following B1937+21 it quickly became clear that many were in binary systems as well. This includes the second known MSP, B1953+29, with a 6.1-ms spin period[9] and 120-day orbital period (Boriakoff et al., 1983), and the third MSP, B1855+09, with a 5.4-ms spin and 12.3-day orbit (Segelstein et al., 1986). The B1855+09 discovery paper notes "The pulsar is only the third one known with P < 10 ms, and the sixth known radio pulsar in a binary system. … Three of the seven binaries are among the fastest five of more than 400 pulsars – a fact that provides strong support for the conclusion that fast pulsars are 'recycled' neutron stars …" B1937+21 being an isolated MSP puts it in the minority and is yet another example of its non-standard nature.

---

[8] Known at the time as *the* binary pulsar; just as B1937+21 was for a time *the* millisecond pulsar.

[9] It is interesting to note that the discovery observation of B1953+29 was taken about 2 years prior to that of B1937+21! However, the spin period was initially misidentified due to insufficient time resolution.

Discoveries continued (see Fig. 5) and these fully-recycled MSP systems eventually became much more numerous than double-neutron star binaries. These "true" MSPs have typical spin periods of a few ms and usually white dwarf companions. Due to the lower companion mass, the duration of mass transfer and spin-up is longer than for double-neutron star systems. This leads to the observed faster spin periods as well as highly circular orbits. This makes mass determination via relativistic orbital precession as was done for B1913+16 challenging or impossible. However, general-relativistic delay of pulses as they pass near a binary companion is measurable and depends only on having a favorable (high) orbital inclination. This effect is known as Shapiro delay and was first measured via solar system radar experiments (Shapiro, 1964). The first demonstration in an MSP system was done using Arecibo observations of B1855+09 (Ryba and Taylor, 1991). The Shapiro delay amplitude is roughly 20 $\mu$s in this case, and is easily seen given the 2 $\mu$s timing uncertainty in these data. The delay is directly proportional to the binary companion mass which, when combined with other measurable orbital parameters, allows inference of the neutron star mass, in this case 1.27$\pm$0.2 solar masses.

The mass of neutron stars, and in particular the maximum observed mass, is of great interest not just for astrophysical reasons but also since it provides a unique constraint on nuclear physics and the properties of ultra-dense matter (e.g. Lattimer and Prakash, 2016). The density and temperature regime present in the star's interior is not easily accessible via laboratory experiments on Earth. Theories predict different equations of state which determine both the radius of a star of given mass, and the maximum mass that can be supported before collapse to a black hole. Even if radius measurement is unavailable, observed high masses can therefore rule out certain theories. Recent MSP observational work in the radio has resulted in several neutron stars that approach or exceed 2 solar masses (Demorest et al., 2010; Antoniadis et al., 2013; Cromartie et al., 2020) Ongoing X-ray measurements from the *NICER* satellite result in both mass and radius measurements which further constrain the nuclear equation of state (e.g. Miller et al., 2019).

Double neutron star binaries such as B1913+16 have typically been the main pulsar targets for tests of gravity due to their high total masses and compact, eccentric orbits. However, the discovery of a pulsar triple system consisting of a neutron star (the pulsar) and two white dwarfs in orbits with periods 1.6 and 327 days provided a unique new gravitational laboratory (Ransom et al., 2014). By analyzing the motion of two unequal-mass objects (the NS and inner WD) in an external gravitational field (from the outer WD), the equivalence principle was verified to orders of magnitude more precision than previously achievable (Archibald et al., 2018).

Despite these impressive results to date, this field remains to some extent sensitivity-limited with potential for additional ground-breaking discoveries to come. It is likely that we have so far only detected ~ 10% of the total galactic population of MSPs that would be accessible with more sensitive radio telescopes (e.g., Keane et al., 2015). This yet-undiscovered population is expected to include novel systems that will provide new tests of gravitation, nuclear physics, and the astrophysics of neutron stars. In particular, a pulsar with a stellar-mass black hole as its binary companion, and a population of MSPs orbiting our galaxy's central supermassive black hole Sgr A* are highly anticipated discoveries. Searches for these form a major part of the science cases for future radio telescopes such as the Square Kilometre

---

Furthermore, the B1953+29 discovery survey targeted unassociated gamma-ray sources detected by the COS-B satellite. In contrast to the "scintar" approach, gamma-ray based target selection eventually became an extremely productive method of MSP discovery via the *Fermi* gamma-ray telescope (e.g. Smith et al., 2023).

Array (SKA, Eatough et al., 2015; Shao et al., 2015; Watts et al., 2015) and the next-generation Very Large Array (ngVLA, Bower et al., 2018).

## 4.2 Pulsar Timing Arrays

One of the most notable, as well as challenging, scientific uses to come out of the discovery of MSPs has been the attempt to use high-precision timing of pulsars as a means of detecting very low-frequency (long period) gravitational waves (GW). This again relies on the basic physical principle of the radio pulses acting as a probe of the intervening media. In this case, GW passing between the pulsar and Earth alter the light travel time between these two points, and affect the pulse times of arrival which are then modulated periodically at the GW frequency. Here the GW are generated by any external source, rather than by the pulsar or its binary system as in the previous section. This concept had been realized and publicized several years prior to the 1982 discovery of MSPs (Sazhin, 1978; Detweiler, 1979). A foundational result in the field was made by Hellings and Downs (1983), who showed that GW signals produce correlated timing fluctuations amongst a set of pulsars. This feature allows GW-induced perturbations to be distinguished from other "timing noise" (e.g., glitches, etc) intrinsic to each pulsar. They derive the specific angular correlation function for the important case of an isotropic GW background (GWB) signal, and use existing data from four canonical pulsars to set an upper limit on such a signal. At the time, a potential cosmological GWB was often considered, and limits were quoted in units of the universe closure density; in these units, the Hellings-Downs limit was $\Omega_{GW} < 1.4 \times 10^{-4}$.

It is interesting to note that despite being published several months after the discovery of B1937+21, no mention of MSPs was made in the Hellings & Downs paper. Nevertheless, the enormous advantages of MSPs for GW detection were soon realized – the faster spin periods and shorter pulses provide orders of magnitude improvement in the amplitude of timing fluctuations that can be measured. Additionally, MSPs are much better clocks than canonical pulsars; being older, and with much smaller magnetic fields, the spin rates of MSPs are much more stable, reducing the level of pulsar-intrinsic timing noise that would mask a small GW signal. Don Backer and collaborators championed this effort during the 1980's and 1990's, and coined the term "pulsar timing array" (now typically abbreviated as PTA) to describe the use of a set, or array, of MSPs for GW detection. The first published use of the term appears in a 1989 Bulletin of the AAS, which reports: "Foster, Backer and Taylor (Princeton) began timing an array of millisecond pulsars. The array approach will allow pulsars to be timed against each other without dependence on the earth clock … Pulsar timing array data will provide the best limits on the cosmic background of gravitational radiation on length scales of a few light years." Foster and Backer (1990) in "Constructing a Pulsar Timing Array" first discuss many of the core concepts and technical challenges and present initial results from an observational program on three MSPs using the NRAO 43-m telescope at Green Bank.

The Princeton pulsar group (Joe Taylor et al.) were important contributors to the PTA effort during this time period, running long-term MSP timing programs at Arecibo (including some results already mentioned in Section 4.1). A key result was published by Kaspi et al. (1994), presenting 8 years of timing data on MSPs B1855+09 and B1937+21. Multiple astrophysical conclusions were derived from these data, one of which was a GWB limit of $\Omega_{GW} < 6 \times 10^{-8}$ based on the B1855+09 data alone. At this point it became clear that despite being one of the brightest and fastest MSPs, B1937+21 suffers from significantly higher levels of timing noise than most others; the reasons for this are still poorly understood, and this is yet another unique feature of the first MSP. Due to its excess noise, B1937+21 would not contribute significantly to future PTA results.

During the late 1990s and early 2000s the Berkeley and Princeton groups collaborated on a "coordinated timing" program at Arecibo of long-term monitoring of MSPs for multiple scientific goals. Using the same observations – sometimes recorded with different backend systems – different analyses would be done for GW, binary properties, ISM studies, etc. This model of relatively large multi-purpose observational programs has continued to the current day in PTA work. From 2005 onwards, interest in PTA work increased significantly, and the approach gradually became recognized as a promising method for GW detection. Several large PTA collaborations developed in this time period: In 2007, the Berkeley-Princeton collaboration expanded, bringing in a number of other pulsar researchers, primarily from institutions in the U.S. and Canada. The new group was formally named the North American Nanohertz

Observatory for Gravitational Waves, or NANOGrav[10]. Parallel efforts were started with the PPTA in Australia (Hobbs et al., 2009) and EPTA in Europe (Janssen et al., 2008). The groups coordinate and share data via the International Pulsar Timing Array (Hobbs et al., 2010).

Currently, pulsar timings array experiments are beginning to detect the first signs of nHz-frequency gravitational waves. The most likely source is now thought to be a stochastic background of unresolved binary supermassive black holes throughout the universe. All such systems with orbital periods on the order of a few years contribute to the signal. In 2020, NANOGrav published analysis showing evidence for a common noise process affecting the timing of 45 MSPs monitored as part of the project for up to 13 years (Arzoumanian et al., 2020). If interpreted as a GW signal the characteristic strain amplitude would be $\sim 2 \times 10^{-15}$. This was followed up recently with an updated data set including over 15 years of data on 68 MSPs (Agazie et al., 2023a). The corresponding GW analysis found significant evidence for the Hellings-Downs correlations that are the signature of the GWB; the strain amplitude was determined to be $2.4^{+0.7}_{-0.6} \times 10^{-15}$ (Agazie et al., 2023b). Contemporaneous results from other PTA collaborations (Antoniadis et al., 2023; Xu et al., 2023; Zic et al., 2023) found consistent results at lower significance levels (The International Pulsar Timing Array Collaboration et al., 2023). Moving forward, PTA work is expected to continue to be a major driver for pulsar observations in the coming decades. After the initial GWB detection, future goals include detailed characterization of the GW spectrum and detection of individual black hole binary systems.

## 5    CONCLUSIONS

The story of PSR B1937+21 was correctly and concisely described by Helfand (1982) as "The story of the discovery is one of perseverance with a dash of good luck." Luck certainly contributed some – the prime example being the chance coincidence on the sky of several unrelated sources. This led both to enough flux density for the original 4C detection to be made, and the incorrect yet highly motivational hypothesis that the source was a young pulsar embedded in a supernova remnant. However the discovery owes far more to the persistence, astrophysical and technical insight, and leadership abilities of Don Backer – in choosing this source as a worthwhile mystery to pursue, recognizing instrumental effects such as the 4C lobe confusion, assembling a capable team who could carry out the various parts of the investigation, and systematically testing different possibilities until arriving at the ultimate answer.

The discovery of millisecond pulsars opened the door to an entire field of study which continues to this day; these future scientific consequences were likely completely unanticipated at the time of the original search. While we may attempt to predict expected future discoveries, we should be sure to not be completely blind to new unexpected possibilties. As noted by Backer (1984), "The discovery of PSR 1937+21 in an unexplored domain of parameter space is a reminder to all scientists that many of nature's best secrets remain to be discovered."

## 6    ACKNOWLEGEMENTS

All personal correspondence cited in this article is from the National Radio Astronomy Observatory / Associated Universities, Inc. Archives, Papers of Donald C. Backer, Radio Astronomy Research Series, Pulsars Unit, PSR 1937+21 Subunit – The authors thank David Backer (Don's son) for permission to publish quotes from this collection. We are also grateful to Heather Cole and Ellen Bouton of the NRAO/AUI Archives for organizing and archiving this material, and to Imke de Pater, Geoff Bower, Dan Werthimer and David DeBoer for their assistance with the initial sorting.

---

[10] http://nanograv.org

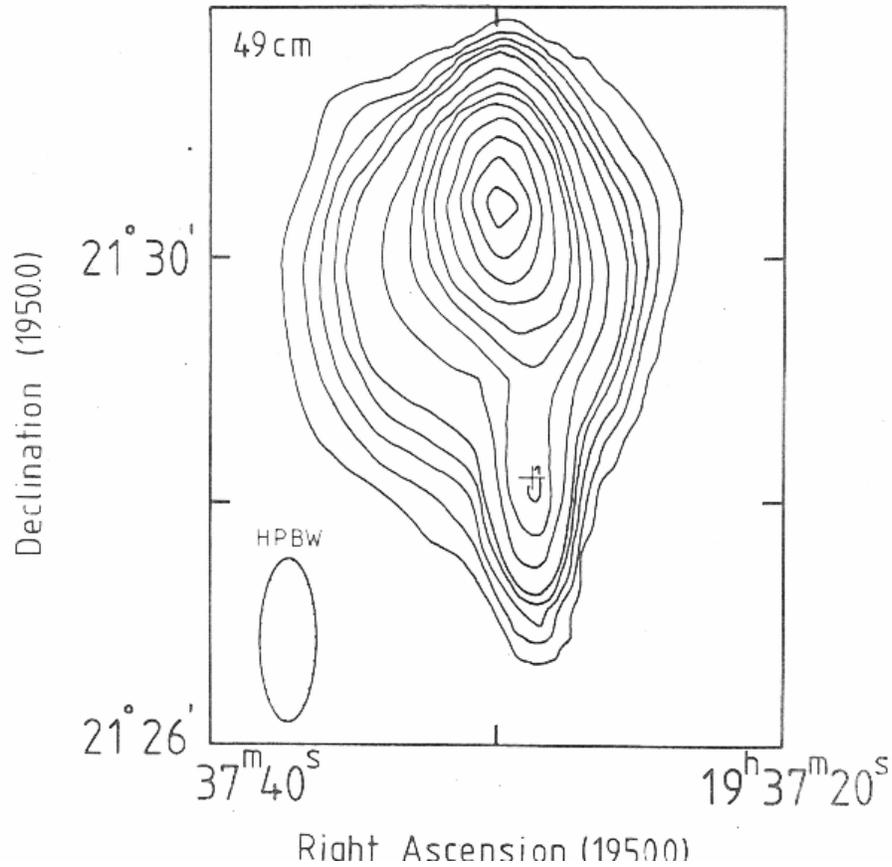

*Figure 1: Image of the PSR 1937+21 region (from the Goss archive). Westerbork Synthesis Radio Telescope (WSRT) at 608.5 MHz or 49 cm. 15 January 1982 with a beam of 31.3 by 80.4 arcsec in RA and dec. Beam is shown in left hand corner. The extended HII region to the north at a distance of about 11 kpc is not related to the southern point source, the pulsar PSR B1937+21 at a distance of about 5 kpc (see Heiles et al., 1983)*

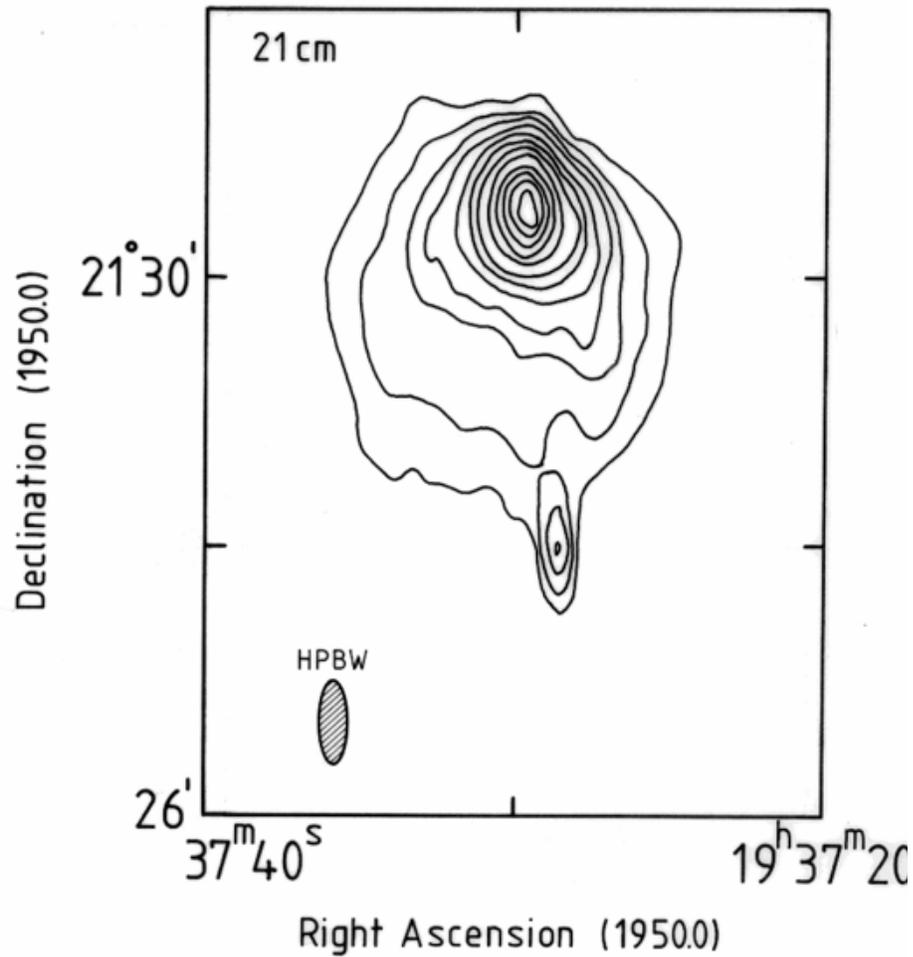

*Figure 2: 20-cm continuum image of the PSR B1937+21 area made with the WSRT (from the Goss archive). The extended HII region to the north and the pulsar to the south. From 8 August 1982, beam is 13.3 by 37 arcsec in RA and dec.*

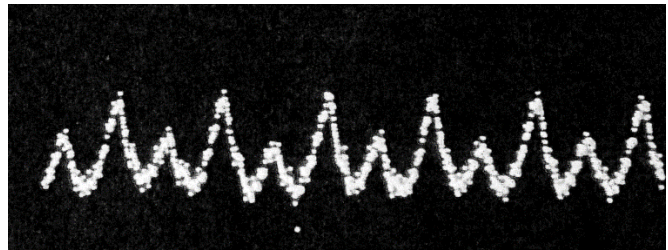

*Figure 3: Detection of PSR B1937+21, radio intensity versus time. The plot begins on the left with the interpulse and spans about 10 ms total, showing six periods of the pulsar (from the NRAO/AUI Archive, Backer Papers; also based on Figure 4 of Backer et al., 1982a).*

### PERIODIC COMET CHURYUMOV-GERASIMENKO (1982f)

Further observations have shown that the orbital solution given on IAUC 3731 is unsatisfactory, with the comet now some 20" west of the ephemeris. The following improved result is based on a total of 69 observations (through 1982 Nov. 8) and yields nongravitational parameters $A_1 = +0.01 \pm 0.09$, $A_2 = +0.0125 \pm 0.0020$. A computation from 36 observations during 1975-1982 alone (and ignoring nongravitational effects) is practically identical.

$$T = 1982 \text{ Nov. } 12.0996 \text{ ET} \qquad \text{Epoch } 1982 \text{ Nov. } 7.0 \text{ ET}$$

| | | |
|---|---|---|
| $\omega = 119.3244$ | | $e = 0.629153$ |
| $\Omega = 50.3592$ ] 1950.0 | | $a = 3.522053$ AU |
| $i = 7.1130$ | | $n° = 0.1491112$ |
| $q = 1.306142$ AU | | $P = 6.610$ years |

M. Wallis, University College, Cardiff, telexes that his observations with IUE on Nov. 7.59 UT gave a fine-error-sensor (blue, central region) mag of 12.4. The corresponding mag for Comet Austin (1982g) was 14.3. On the other hand, the ultraviolet spectra showed that comet 1982f was less bright in OH than comet 1982g by one magnitude, yet unlike comet 1982g had continuum and molecular ultraviolet emissions.

### MILLISECOND PULSAR IN 4C 21.53

D. Backer, S. Kulkarni and C. Heiles, University of California at Berkeley; M. Davis, Arecibo Ionospheric Observatory; and M. Goss, University of Groningen, report the detection of a millisecond pulsar in 4C 21.53. The period was 0.001557708 s on Sept. 25. The period derivative is $3 \cdot 10^{-14}$, and the dispersion measure is 100 e pc cm$^{-3}$. Arecibo observations at 1400 MHz show deep fading from iss over 5 MHz and 10 min. The position from VLA observations is $\alpha = 19^h 37^m 28^s.72$, $\delta = +21°28'01".3$ (equinox 1950.0).

### Z ANDROMEDAE

With reference to the item on IAUC 3738, J. Bortle, Stormville, NY, remarks that this star does not seem to have had a recent photometric outburst. His visual magnitude estimates follow: June 15.1 UT, 10.8; July 22.1, 10.9; Aug. 17.1, 10.8; Sept. 10.1, 10.8; Oct. 11.1, 10.9; Oct. 22.0, 10.8; Nov. 3.0, 10.8.

1982 November 12                                         Brian G. Marsden

*Figure 4: IAU Circular No. 3743 reporting the initial detection of B1937+21 as a millisecond pulsar (courtesy Daniel W. E. Green and the Central Bureau for Astronomical Telegrams).*

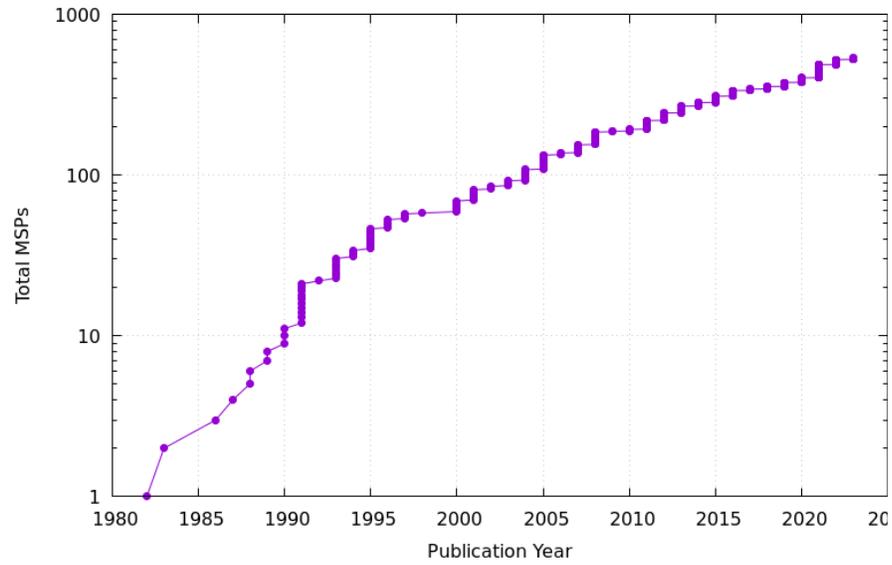

*Figure 5: Cumulative number of known MSPs (defined here as those having P < 20 ms) as a function of publication year. Data from the ATNF Pulsar Catalog v1.70. (Manchester et al., 2005, http://www.atnf.csiro.au/research/pulsar/psrcat)*